# Parallel Betweenness Computation in Graph Database for Contingency Selection

Yongli Zhu, Renchang Dai, Guangyi Liu

GEIRI North America, San Jose, USA (yongli.zhu@geirina.net, renchang.dai@geirina.net, guangyi.liu@geirina.net)

*Abstract*—Parallel betweenness computation algorithms are proposed and implemented in a graph database for power system contingency selection. Principles of the graph database and graph computing are investigated for both node and edge betweenness computation. Experiments on the 118-bus system and a real power system show that speed-up can be achieved for both node and edge betweenness computation while the speeding effect on the latter is more remarkable due to the data retrieving advantages of the graph database on the power network data.

## I. INTRODUCTION

To date, in the operation and planning of modern power systems, contingency analysis is a vital task that can provide critical information to system operators and planners regarding the fragility of the system under the loss of components. The loss of components can be either unplanned like a hurricane event or planned like pre-scheduled maintenance of power lines. N-1 contingency analysis is widely utilized in system operation as a common practice, where a component (e.g. either a bus or a line) is removed and the power flow procedure is performed on the remaining system to see if any line flows or bus voltages will violate the pre-defined security criteria.

Among different strategies in the selection of critical lines or buses for contingency analysis (in short, *contingency selection*), the betweenness values of the nodes and edges can provide insights from the structural viewpoint and have attracted more and more attention from different researchers. In [1], a novel Performance Index (PI) by combining the closeness centrality and betweenness centrality is proposed for contingency ranking in the N−x (x=1,2,3) analysis. In [2], the concept of weighted-line-betweenness is proposed to identify the vulnerable lines in a power grid. Similarly, paper [3] also adopts the betweenness concept. By using the reactance of the transmission line as the weight, it defines a new betweenness index as the electric path from one bus to another.

However, one challenge in applying the betweenness in real power system application is its computing speed. The time complexity of the naive algorithm is about $O(|V|^3)$, where $|V|$ is the total number of nodes in the graph; the most cutting-edge algorithm, i.e. the Brandes algorithm is about $O(|V||E|)$. For a power grid with thousands of buses and branches, that time complexity can result in a huge computational burden in the conventional serial computing paradigm. In addition, the space complexity is also a big issue. Conventionally, a power network is represented in matrix (2D) format in the computer. The matrix itself can be very large but sparse, which makes it difficult to be stored entirely in the computer memory, but in the hard disk or database. However, on one hand, a large portion of memory will be wasted in this way; on the other hand, the read/write efficiency can be extremely low. To overcome this challenge, the graph database is utilized [4].

In this paper, a novel graph computing-based parallel algorithm for betweenane calculation is presented and implemented in a graph database. Section II introduces the basics of node betweenness and edge betweenness together with the existed serial algorithms. Section III briefly explains the programming model of the graph database. Section IV presents the proposed parallel betweenness algorithms implemented in the graph database. Section V presents the case studies on the IEEE 118-bus system and a real province power system of China. Concluding remarks are given in the final section.

## II. BETWEENNESS AND SERIAL ALGORITHM

### A. Node Betweenness

In graph theory, betweenness centrality (in short, **betweenness**) is a measure of centrality in a graph based on shortest path concepts. For every pair of nodes in a connected graph, there exists at least one shortest path between two nodes such that either the number of edges that lie on this path (for unweighted graphs) or the sum of the weights of those edges (for weighted graphs) is minimized. The betweenness centrality of a node $v$ is the sum of the fraction of all-pairs shortest paths that pass $v$, which is given by Eq. (1):

$$C_B(v) = \sum_{s \neq v \neq t} \frac{\sigma_{st}(v)}{\sigma_{st}} \qquad (1)$$

where $\sigma_{st}$ is the total number of shortest paths from node $s$ to node $t$ and $\sigma_{st}(v)$ is the number of those paths that passes $v$ (excluding $v$ as the two ends). Normalization can be also executed, i.e. dividing the above formula by $(N-1)(N-2)/2$.

This work was supported by China State Grid Corporation technology project 5455HJ180018

The node betweenness centrality has played an important role in network theory. It reflects the degree to which nodes exist between each other. A node with higher betweenness centrality may have more influence over the network because more information/energy flow will pass through that node.

### B. Edge Betweenness

The edge betweenness centrality of an edge *e* is the sum of the fraction of all-pairs shortest paths that passes *e*:

$$C_B(e) = \sum_{s,t \in V} \frac{\sigma_{st}(e)}{\sigma_{st}} \quad (2)$$

where, $\sigma_{st}(e)$ is the number of those paths that pass through *e*. Note that in definition (2), here nodes *s* and *t* can be identical. The edge betweenness can be used for the selection of most critical lines in a power grid.

### C. Serial Algorithms for Betweenness Calculation

#### 1) Brandes Algorithm for Vertex Betweenness

The *Brandes* Algorithm [5] is the most famous algorithm specially designed for betweenness calculation. To compute the betweenness for all the nodes in a graph **G** = <**V**, **E**>, where **V** is the node-set, **E** is the edge set, its time complexity can be reduced to $O(|V|*|E|)$ compared to the naïve algorithm in $O(|V|^3)$. The space complexity is about $O(|V|+|E|)$.

First, defines the *pair-dependency* $\delta_{st}(v)$ as follows:

$$\delta_{st}(v) = \frac{\sigma_{st}(v)}{\sigma_{st}} \quad (3)$$

Thus, Eq. (1) becomes:

$$C_B(v) = \sum_{s \neq v \neq t} \delta_{st}(v) = \sum_{s \neq v} \delta_s(v) \quad (4)$$

where,

$$\delta_s(v) = \sum_{t \neq v} \delta_{st}(v) \quad (5)$$

It can be proofed that the following equation holds [5]:

$$\delta_s(v) = \sum_{w:v \in P_s(w)} \frac{\sigma_{sv}}{\sigma_{sw}}(1 + \delta_s(w)) \quad (6)$$

$\delta_s(v)$ is the pair-dependency value of node *v* with starting node *s*. Finally, the betweenness centrality of node *v* is nothing but the sum of all pair dependency values from all node *s* ($\neq v$). $P_s(w)$ is all the precedent nodes of *w*. The pseudocode is shown as follows [5].

| Algorithm 1: Node Betweenness (serial computing) |
|---|
| 1 **Input**: **G** = <**V**, **E** >, number of vertices *n* |
| 2 $C_B[v] \leftarrow 0, \forall v \in V$ |
| 3 **for** $s \in V$ |
| 4     $S \leftarrow$ empty stack |
| 5     $P[w] \leftarrow$ empty list, $w \in V$ |
| 6     $\sigma[t] \leftarrow 0, t \in V; \sigma[t] \leftarrow 1$ |
| 7     $d[t] \leftarrow -1, t \in V; d[s] \leftarrow 0$ |
| 8     $Q \leftarrow$ empty queue |
| 9     enqueue $s \in Q$ |
| 10    **while** *Q* not empty |
| 11      dequeue $v \leftarrow Q$ |
| 12      push $v \to S$ |
| 13      **for** neighbor *w* of *v* |
| 14       **if** $d[w] < 0$   // *w* is found for the first time |
| 15        enqueue $w \to Q$ |
| 16        $d[w] \leftarrow d[v] + 1$ |
| 17       **if** $d[w] = d[v] + 1$ //shortest path from *s* to *w* via *v* |
| 18        $\sigma[w] \leftarrow \sigma[w] + \sigma[v]$ |
| 19        append $v \to P[w]$ |
| 20    $\delta[v] \leftarrow 0, v \in V$ |
| 21    // *S* pop vertices in descending order of distance from *s* |
| 22    **while** *S* not empty |
| 23      pop $w \leftarrow S$ |
| 24      **for** $v \in P[w]$ |
| 25       $\delta[v] \leftarrow \delta[v] + (1 + \delta[w]) \cdot \sigma[v]/\sigma[w]$ |
| 26      **if** $w \neq s$ |
| 27       $C_B[w] \leftarrow C_B[w] + \delta[w]$ |
| 28 **Output**: a table of betweenness values for all the vertices |

#### 2) Brandes Algorithm for Edge Betweenness

The *Brandes* Algorithm can be also adapted to compute edge betweenness in the same level of time complexity. Its principle is based on that of node betweenness computation, i.e. fixing a source node at the beginning of each high-level literation, then calculating a new value of path dependency for each pair of the source node and midway node, finally accumulating it to the corresponding edge.

The only code which needs adaption is the part from line 22 of the above code as shown in the following pseudocode.

| Algorithm 2: Edge Betweenness (serial computing) |
|---|
| 1 **Input**: **G** = <**V**, **E** >, number of vertices *n* |
| …… |
| 22    **while** *S* not empty |
| 23      pop $w \leftarrow S$ |
| 24      **for** $v \in P[w]$ |
| 25       $temp \leftarrow (1 + \delta[w]) \cdot \sigma[v]/\sigma[w]$ |
| 26       $C_B[(v,w)] \leftarrow C_B[(v,w)] + temp$ |
| 27       $\delta[v] \leftarrow \delta[v] + temp$ |
| 28      **if** $w \neq s$ |
| 29       $C_B[w] \leftarrow C_B[w] + \delta[w]$ |
| 30 **Output**: a table of betweenness values for all the edges |

## III. GRAPH DATABASE AND GRAPH COMPUTING

### A. Graph Database

Traditional *Relational Database* like MySQL is based on the relational model of data, and its basic operations like create, read, update, and delete (CRUD) are based on SQL (Structured Query Language). However, with the emerging of various new business models like social networks, which may have millions of "users" and their associated "relationships", the operation speed of the relational database may deteriorate drastically. Therefore, non-relational databases that store real data in the forms of "edge" and "vertices" with higher read/write efficiencies are given birth, e.g. graph database [6].

*B. Graph Computing*

Here, "graph computing" means "computing (either numerical or non-numerical) implemented in the graph database". The basic idea of graph computing can be summarized as three steps: "Communication", "Synchronization" and "Computation". After the original graph is divided and dispersed among different "Workers" (e.g. threads or cores), the following procedures will happen within each "Super-step" (i.e. one high-level iteration):

- **Communicating** any necessary messages among different vertices and two consecutive "Super-steps"
- **Synchronization** of all intermediate results/messages from the vertices in the previous "Super-step"
- **Computation** is executed by all the "Workers" in the current "Super-step" and execution orders are random.

Thus, graph computing provides a novel parallel *programming paradigm* that is suitable for problems with inner graph structures like social networks and power grids.

## IV. PARALLELIZATION OF BETWEENNESS ALGORITHMS IN GRAPH DATABASE

The graph database provides a native platform for parallelizing most graph theory related algorithms, especially for Breath-First-Search (BFS) based algorithms such as bridge detection [4], connected component search, etc. In this section, the parallel algorithms for both the node and edge betweenness computation are implemented in the graph database. The core idea is to disperse the part of the shorted-path computation of each pair of nodes into different "Workers" via graph database.

The pseudocode for computing the node and edge betweenness in the graph database is shown as follows, where "**accum**" and "**while**" are the built-in keywords of the graph database query language, meaning that the "for/while-loop" computation will be *implicitly* dispersed to parallel "Workers" and be executed for all queried nodes or edges. In the graph database, each computation task is carried by a specific *query* (similar to the *function* concept in serial programming).

*A. Parallel Node-Btw Computing in Graph Database*

The basic idea is based on BFS. The pseudocode consists of two parts: the main query and the sub-query (similar to the relationship between "main function" and its "sub-function" in the conventional serial programming paradigm).

---

**Algorithm 3: Main query for node-betweenness (parallel computing)**

1 **Input**: $G = <V, E>$, number of nodes $n$
2 int nodeNum = 1
3 Start = {All Nodes} // A set of all nodes
4 nodeNum = Start.size()
5 _ = **select** $s$ **from** Start  // $\forall s \in$ Start
6     **accum**  // for each source node $s$
7         **nodebtw_subquery** ($s$, nodeNum)
8 **Output**: an output file of all the node betweenness values

---

**Algorithm 4: Sub-query for node-betweenness (parallel computing)**

1 **nodebtw_subquery** ($s$, nodeNum) // $v$ is the source node
2 Start = {$s$}  // Starting set initialization
3 $s$.@dist = 0 // distant to itself is zero
4 $s$.@spNum = 1 // only one path to itself
5 **while** Start.size() > 0
6   temp = Start
7   Start = **select** $t$ **from** all $e = (v, t)$  $\forall v \in$ Start
8     **where** $t$.@dist < 0
9     **accum**
10       $t$.@dist = $v$.@dist + 1
11   _ = **select** $t$ **from** all $e = (v, t)$  $\forall t \in$ temp
12     **where** $t$.@dist >= 0 and $t$.@dist == $v$.@dist + 1
13     **accum**
14       $t$.@spNum += $v$.@ spNum
15   @@currDist += 1
16 **if** Start.size() == 0
17   @@currDist += −1
18   Start = temp
19 **while** @@currDist > 0
20   Start = **select** $v$ **from** all $e = (v, t)$  $\forall v \in$ Start
21     **where** $t$.@dist > $v$.@dist
22     **accum**
23       $v$.@pd += $v$.@spNum*1.0 /$t$.@spNum*(1+$t$.@pd)
24     **post-accum**
25       **if** $v$.Id != $s$.Id then
26         $v$.btw += $v$.@pd
27   Start = **select** $t$ **from** V    // $\forall t \in$ V
28     **where** $t$.@dist == @@currDist − 1
29   @@currDist += −1

---

In the above code, @dist is a node property attached on each node entity in the graph database, e.g. $v$.@dist represents the shortest distance from the source node $s$ to $v$. Similarly, $v$.@spNum stands for the total number of the shortest paths starting from source node $s$ and passing by the node $v$. $v$.@pd represents the "pair dependency" value from the source node $s$ to $v$, i.e. $\delta_s(v)$ in Eq. (5).

The first while loop begins with the given source node $s$ and calculates 1) the number of shortest paths for each node that has been passed by 2) the shortest path length for each node (starting from source node $s$). Thus, it can be named as the "forward while-loop". @@currDist is the counter to record the so-far found distance value of the "shortest path" during the "forward while-loop".

The second while-loop is used for updating those "pair-dependency" values for all the nodes from the farthest level of nodes. Thus, it can be named as the "backward while-loop". Eq. (2) is involved in this while-loop. Note that, 1) the condition "$t$.@dist > $v$.@dist" is to guarantee that the pair-dependency is updated from the "farthest" (the largest value of shortest paths) to the source node $s$; 2) the condition " where

$t.@\text{dist} == @@\text{currDist} - 1$" is used to select the next nearest group of the nodes for the backward updating of the pair-dependency values.

Fig. 1 and Fig. 2 illustrate one intermediate step of the above parallel node betweenness computation algorithm.

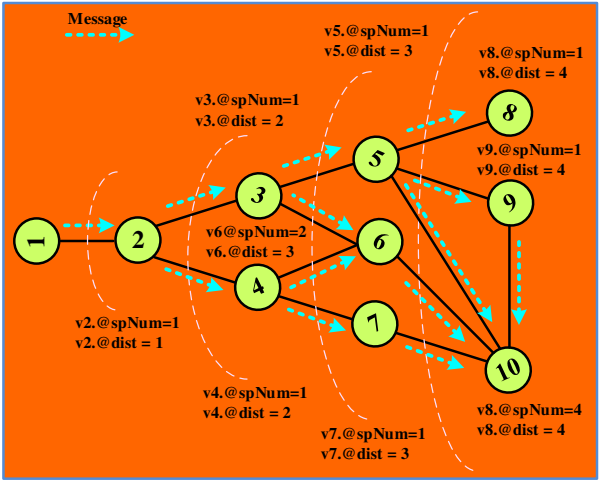

Figure 1. Illustration of Graph Computing based Node Betweenness Calculation: the forward while-loop

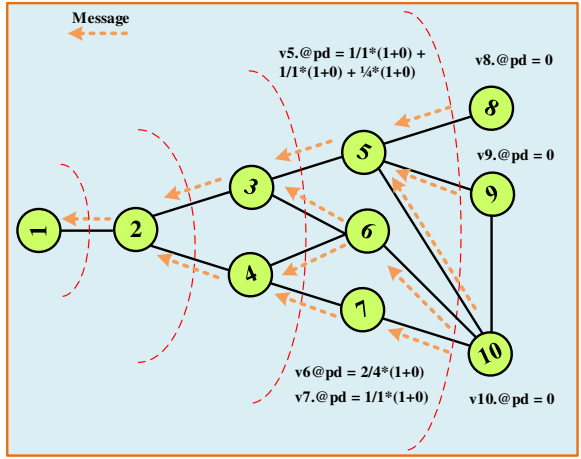

Figure 2. Illustration of Graph Computing based Node Betweenness Calculation: the backward while-loop

### B. Parallel Edge-Btw Computing in Graph Database

**Algorithm 5: Main query for edge-betweenness (parallel computing)**

1  **Input**: $G = <V, E>$, number of vertices $n$
2  int nodeNum = 1
3  Start = {All Nodes}
4  nodeNum = Start.size()
5  _ = **select** $s$ **from** Start  // $\forall s \in$ Start
6      **accum**
7          **edgebtw_subquery** ($s$, nodeNum)
8  **Output**: an output file of all the edge betweenness values

For edge betweenness computation, its pseudocode is shown in "**Algorithm 5**" and "**Algorithm 6**", where "$e.\text{btw}$" is an edge property representing the edge-betweenness value.

**Algorithm 6: Sub-query for edge-betweenness (parallel computing)**

1  **edgebtw_subquery** ($s$, nodeNum)
2  Start = {$s$}   // Starting set initialization
3  $s.@\text{dist} = 0$ // distant to itself is zero
4  $s.@\text{spNum} = 1$ // only one path to itself
5  **while** Start.size() > 0
6      temp = Start
7      Start = **select** $t$ **from** all $e = (v, t)$ $\forall v \in$ Start
8          **where** $t.@\text{dist} < 0$
9          **accum**
10             $t.@\text{dist} = v.@\text{dist} + 1$
11     _ = **select** $t$ **from** all $e = (v, t)$ $\forall t \in$ temp
12         **where** $t.@\text{dist} >= 0$ and $t.@\text{dist} == v.@\text{dist} + 1$
13         **accum**
14             $t.@\text{spNum} += v.@\text{spNum}$
15     @@currDist += 1
16 **if** Start.size() == 0
17     @@currDist += −1
18     Start = temp
19 **while** @@currDist > 0
20     Start = **select** $v$ **from** all $e = (v, t)$ $\forall v \in$ Start
21         **where** $t.@\text{dist} > v.@\text{dist}$
22         **accum**
23             $v.@\text{pd} += v.@\text{spNum}*1.0 \,/t.@\text{spNum}*(1+t.@\text{pd})$
24             $e.\text{btw} += v.@\text{spNum}*1.0 \,/t.@\text{spNum}*(1+t.@\text{pd})$
25         **post-accum**
26             **if** $v.\text{Id} \,!= s.\text{Id}$ then
27                 $v.\text{btw} += v.@\text{pd}$
28     Start = **select** $t$ **from** V    // $\forall t \in$ V
29         **where** $t.@\text{dist} == @@\text{currDist} - 1$
30     @@currDist += −1

## V. CASE STUDY

In this section, the IEEE 118-bus system and China Sichuan Province system are tested for the proposed algorithms (without normalization). All the experiments are implemented on a Linux server with 88 Intel Xeon CPU E5-2699A v4 @ 2.40GHz and 128GB memory. The default number of threads is 8.

### A. IEEE 118-bus system

The 118-bus system has 118 buses and 186 branches [7].

#### a) Node betweenness result

TABLE I. NODE BETWEENNESS FOR 118-BUS SYSTEM

|          | Parallel computing | Serial computing |
|----------|--------------------|------------------|
| Time (ms)| 141.596            | 169.655          |
| Node #   | Btw value          | Btw value        |
| 1        | 1.8333             | 1.8333           |
| 2        | 17.9860            | 17.9860          |
| 3        | 105.6805           | 105.6807         |

#### b) Edge betweenness result

TABLE II. EDGE BETWEENNESS FOR 118-BUS SYSTEM

|  | Parallel computing | Serial computing |
|---|---|---|
| Time (ms) | 136.993 | 233.770 |
| Edge # (from-to) | Btw value | Btw value |
| 1-2 | 20.8194 | 20.8193 |
| 1-3 | 99.8475 | 99.8473 |
| 2-12 | 132.1525 | 132.1527 |

### B. Sichuan system

The China Sichuan power system has 2749 busses and 3280 branches [8]. Its single line diagram is shown in Fig. 3.

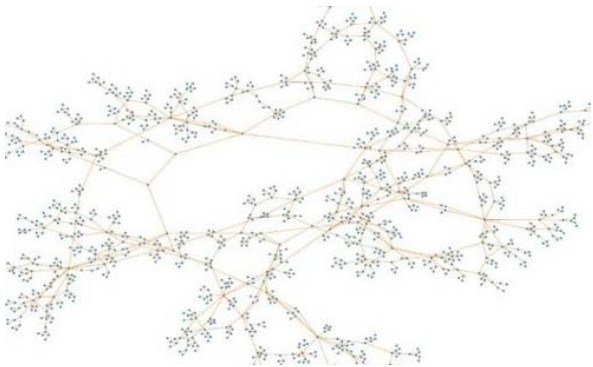

Figure 3. The partial single line diagram of the Sichuan power system

#### a) Node betweenness result

TABLE III. NODE BETWEENNESS FOR SICHUAN POWER SYSTEM

|  | Parallel computing | Serial computing |
|---|---|---|
| Time (ms) | 22943.13 | 85064.89 |
| Node # | Btw value | Btw value |
| 1 | 5493 | 5493 |
| 787 | 13723 | 13723 |
| 1282 | 10982 | 10982 |

#### b) Edge betweenness result

TABLE IV. EDGE BETWEENNESS FOR SICHUAN POWER SYSTEM

|  | Parallel computing | Serial computing |
|---|---|---|
| Time (ms) | 30635.09 | 478416.17 |
| Edge # (from-to) | Btw value | Btw value |
| 1-62 | 2748 | 2748 |
| 1-896 | 2748 | 2748 |
| 1-324 | 8238 | 8238 |

It can be observed that the time cost of edge-betweenness by graph computing is far less than that by conventional serial computing. The reasons are not only due to the parallel nature of the proposed algorithms but also due to the superiority of the graph database in data-retrieving, i.e. storing large sparse matrix (2-D) data in the *node-edge* format to reduce overhead cost in reading/writing memory.

### C. Effects of Different Thread Numbers

To investigate the effect of the number of threads on the speed performance, four different thread numbers are tested on the Sichuan power system, i.e. 4, 8, 16, 32. As shown in Table. V, the best results are obtained at 16 and 8 threads respectively for node and edge betweenness computation. The larger number of threads might lead to larger overhead communication cost among different cores, which can result in slower performance.

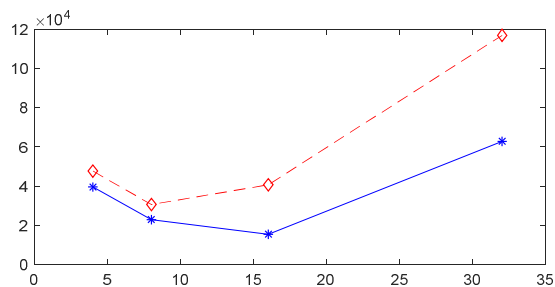

Figure 4. Effects of different thread numbers on the speed performance

TABLE V. EFFECTS OF DIFFERENT THREAD NUMBERS

| Threads | 4 | 8 | 16 | 32 |
|---|---|---|---|---|
| Node | 39548.16 | 22943.13 | **15476.77** | 62826.51 |
| Edge | 47660.11 | **30635.09** | 40694.04 | 116879.3 |

## VI. CONCLUSION

In this paper, graph computing algorithms for betweenness calculation are proposed and implemented in a graph database. Results demonstrate the faster speed of graph computing compared to conventional serial computing on this specific problem. The next step is to deploy this algorithm in a real EMS platform within the contingency selection module based on the betweenness rankings.

ACKNOWLEDGMENT

The authors would like to sincerely appreciate the support and discussion provided by the TigerGraph company.

REFERENCES

[1] A.K. Srivastava, T.A. Ernster, R. Liu and V.G. Krishnan, "Graph-theoretic algorithms for cyber-physical vulnerability analysis of power grid with incomplete information," J. Mod. Power Syst. Clean Energy vol.6, no.5, pp. 887-899, 2018.
[2] X. Chen, K. Sun, Y. Cao and S. Wang, "Identification of Vulnerable Lines in Power Grid Based on Complex Network Theory," IEEE Power Engineering Society General Meeting, Tampa, FL, 2007, pp. 1-6.
[3] A. Dwivedi, X. Yu and P. Sokolowski, "Identifying vulnerable lines in a power network using complex network theory," IEEE International Symposium on Industrial Electronics, Seoul, 2009, pp. 18-23.
[4] Y. Zhu, L. Shi, R. Dai and G. Liu. Fast Grid Splitting Detection for N-1 Contingency Analysis by Graph Computing. The 2019 IEEE PES ISGT-ASIA, Chengdu, China, May.21-24, 2019.
[5] U. Brandes, "A faster algorithm for betweenness centrality," The Journal of Mathematical Sociology, vol. 25, no. 2, pp. 163-177, 2001.
[6] TigerGraph. https://www.tigergraph.com/.
[7] L. Wei, A. I. Sarwat, W. Saad and S. Biswas, "Stochastic Games for Power Grid Protection Against Coordinated Cyber-Physical Attacks," *IEEE Trans. Smart Grid*, vol. 9, no. 2, pp. 684-694, 2018.
[8] R. Dai, X. Zhang, J. Shi, G. Liu, C. Yuan Z. Wang, "Simplify Power Flow Calculation Using Terminal Circuit and PMU Measurements," 2019 IEEE Power & Energy Society General Meeting (PESGM), Atlanta, GA, USA, 2019, pp. 1-5.